\documentclass[letterpaper, 10 pt, journal, twoside]{ieeetran}
\IEEEoverridecommandlockouts
\usepackage{amsmath,amssymb,amsfonts}
\usepackage{algorithmic}
\usepackage{graphicx}
\usepackage{textcomp}
\usepackage{amsthm}
\usepackage{cite}

\usepackage[linesnumbered,ruled,vlined,algo2e]{algorithm2e}%for the environment

\newtheorem{definition}{Definition}

\newcommand{\bt}{\mathcal{T}}

\usepackage{xcolor}
\def\BibTeX{{\rm B\kern-.05em{\sc i\kern-.025em b}\kern-.08em
    T\kern-.1667em\lower.7ex\hbox{E}\kern-.125emX}}
\begin{document}

\title{Improving the Performance of Backward Chained \\ Behavior Trees that use Reinforcement Learning
}
\author{
   Mart Kartašev, %\thanks{Robotics, Perception and Learning Lab., School of Electrical Engineering and Computer Science, Royal Institute of Technology (KTH), SE-100 44 Stockholm, Sweden}
    \and 
    Justin Salér,
    \and
    %XY \and % \thanks{Advanced Concepts Team, European Space Agency, Noordwijk ann Zee, The Netherlands}
    and  Petter \"Ogren 
%    \thanks{Manuscript received: February, 22nd, 2020; Revised May, 18th, 2020; Accepted July, 5th, 2020.}%Use only for final RAL version
%\thanks{This letter was recommended for publication by the Editor Tamim Asfour upon evaluation of the Associate Editor and Reviewers' comments. This  work  was  supported  by SSF  through  the  Swedish  Maritime Robotics Centre (SMaRC) (IRC15-0046), and FOI project 7135.}%Use only for final RAL version
    \thanks{The authors are with the Robotics, Perception and Learning Lab., School of Electrical Engineering and Computer Science, Royal Institute of Technology (KTH), SE-100 44 Stockholm, Sweden, {\tt\small kartasev@kth.se}}
    \thanks{Digital Object Identifier (DOI): see top of this page.}
}

\maketitle

%\markboth{Submitted to IEEE Robotics and Automation Letters. }
%{Kartasev \MakeLowercase{\textit{et al.}}: Improving Behavior Trees using Reinforcement Learning} 
% Use only for final RAL version

\begin{abstract}
In this letter we show how to improve the performance of backward chained behavior trees (BTs) that use reinforcement learning (RL).
BTs represent a hierarchical and modular way of combining control policies into higher level control policies. Backward chaining is a design principle for the construction of BTs that combine reactivity with goal directed actions in a structured way.
The backward chained structure has also enabled convergence proofs for BTs, identifying a set of local conditions that lead to the convergence of all trajectories to a set of desired goal states. 

The key idea of this letter is to improve performance of backward chained BTs by
using the conditions identified in a theoretical convergence proof to setup the RL problems for individual controllers.
In particular, previous analysis identified so-called active constraint conditions (ACCs), that should not be broken in order to avoid having to return to work on previously achieved subgoals.
We propose a way to setup the RL problems, such that they do not only achieve each immediate subgoal, but also avoid violating the identified ACCs.
The resulting performance improvement depends on how often ACC violations occurred before the change, and how much effort was needed to re-achieve them. 
The proposed approach is illustrated in a dynamic simulation environment.
\end{abstract}

\begin{IEEEkeywords}
Behavior trees, Reinforcement learning, Autonomous systems, Artificial Intelligence
\end{IEEEkeywords}

\section{Introduction }

\IEEEPARstart{B}{ehavior trees} (BTs) 
are a modular and hierarchical tool for combining a set of control policies into more complex high level control policies. Reinforcement learning (RL) on the other hand, is a way to create control policies based on reward. Even though many complex problems have been solved using only RL, so-called end-to-end learning \cite{vinyals2019grandmaster}, it is reasonable to believe that in the coming years, many complex robotics problems will still need to be solved by a combination of policies tailored for different sub-tasks. A BT is a way to combine such policies and it is therefore important to investigate how BTs can be combined with RL.

Robot control systems are often quite complex, and modularity is a well known tool for handling complexity.
Ideally, different modules can be developed and tested separately, and then put together to form a bigger system.
Hierarchical modularity, where a module is made up of submodules, is also beneficial, since having a single layer of modules in a big system either leads to very large modules, or a very large set of modules.
BTs are a hierarchically modular tool for composing control policies. In fact, BTs have been shown to be an optimally modular decision structure \cite{biggar2020modularity}. Modularity was also one of the primary reasons for the initial development of BTs in the computer game industry \cite{mateas2003faccade}. 

\begin{figure}[!t]
    \centering
    \includegraphics[width=\columnwidth]{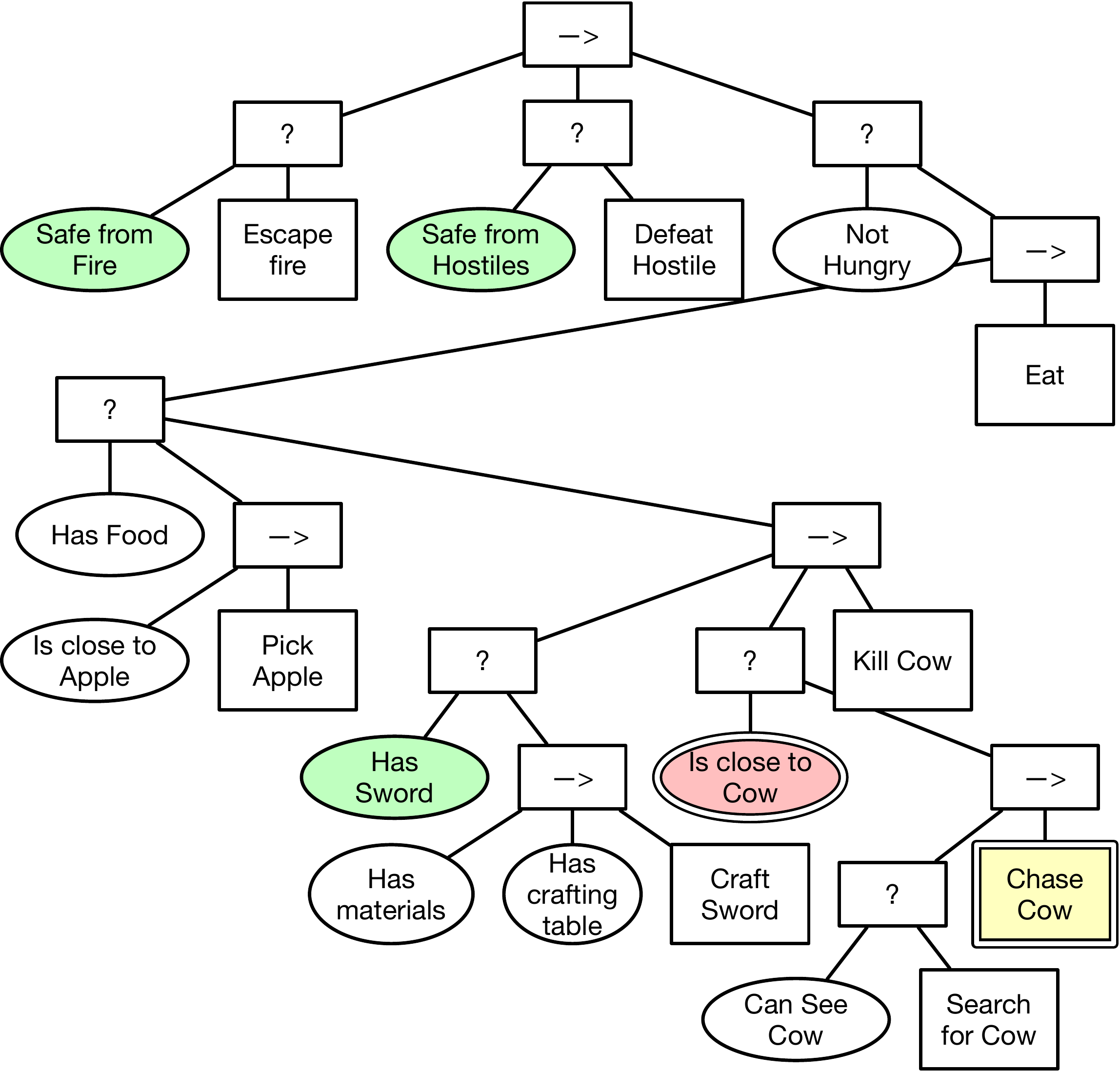}
    \caption{A backward chained BT, the result of recursively replacing conditions with small BTs achieving those conditions, as illustrated in Figure~\ref{fig:PPA_connect}. The ACC concept is illustrated for the action \emph{Chase Cow} (yellow double rectangles). 
    The convergence proof of \cite{ogren2020convergence} states that all actions need to achieve their postconditions, while not violating their so-called ACCs. Thus,
    when using RL to improve this action, there should be a positive award for achieving the postcondition \emph{Is close to cow} (red double ovals) but also a negative reward for violating the ACCs, \emph{Safe from fire, safe from hostiles and has sword} (green ovals).}
    \label{fig:ACC_full_BT}
\end{figure}

In this letter we will show how to use the structure of so-called backward chained BTs, see Figure~\ref{fig:ACC_full_BT}, to implement a controller using RL.
Ideally, there are few dependencies between different parts of a system, but if a particular submodule reverses the achievements of previous submodules, the overall system is not likely to work well. Thus there is a need to identify the dependencies between different modules and create design specifications out of these dependencies to improve performance. We make use of previously published theoretical convergence proofs \cite{ogren2020convergence} to identifying such dependencies, and use them to set up RL problems with consideration for earlier achievements.

To be specific, consider the simple example in Figure~\ref{fig:ACC_full_BT}.
The overall goal of the agent is to be safe from fire, safe from hostiles and not hungry. 
\footnote{In search of an openly accessible simulation environment that is highly complex,  in terms of having many independent agents with different objectives, many different types of objects, and many ways to interact with both objects and agents, we have chosen Project Malm\"o, created by Microsoft Research \url{https://www.microsoft.com/en-us/research/project/project-malmo/}.}
In order to satisfy the final objective, it is currently chasing a cow. The local goal is to get close to the cow in order to kill it. However, the convergence analysis in \cite{ogren2020convergence} shows that there are also a set of non-local goals that needs to be preserved. This includes not stepping into fire, staying away from hostiles, and keeping the sword that is later needed to kill the cow. If any of these are violated, the agent needs to stop chasing the cow and make sure they are satisfied before starting the chase again.
Similar non-local constraints can be identified for any node, in any backward chained BT.

The main contribution of this letter is an approach for combining RL with backward chained BTs in a structured way, based on the convergence proof in \cite{ogren2020convergence}. 
This is done by using the prerequisites of the proof as positive and negative rewards in the RL formulation.
Thus the learning process will not only take into account the natural local subgoals, but also other, non-local constraints that are vital for the progress and performance of the overall policy, of which the RL problem is only a small part. 

The outline of this letter is as follows.
In Section~\ref{sec:related_work} we describe the related work, and 
in Section~\ref{sec:background} we give a background on RL and BTs. 
Then the proposed approach is presented in Section~\ref{sec:proposed_approach},
followed by experiments in Section \ref{sec:experimental_results}.
Finally, conclusions can be found in Section~\ref{sec:conclusions}.

\section{Related Work}
\label{sec:related_work}

In this section we will describe earlier work investigating different aspects of combinations of learning with BTs \cite{pereira2015framework,dey2013ql,fu2016reinforcement,hannaford2016simulation,sprague2018adding,zhang2017combining}. 
There are a number of different ways of incorporating RL into a BT. One can either replace or design a single leaf node using RL, as was done in \cite{pereira2015framework}, or one can replace the interior composition nodes, sequence and fallback, with an RL node where the actions are the children of the original node, as was done in \cite{dey2013ql,fu2016reinforcement,hannaford2016simulation,zhang2017combining}. One can also apply learning on the tree structure itself, moving, adding and deleting subtrees to improve the performance of the overall policy 
\cite{lim2010evolving,nicolau2016evolutionary,colledanchise2018learning,paduraru2019automatic}.

In \cite{dey2013ql} the authors aim to improve an existing BT. Given a reward function,  the lowest level sequences of the BT are identified and used as actions in an RL problem. The q-value is estimated, and used to create so-called q-conditions for each action, returning success when the corresponding action has the highest q-value. Using these, a BT is constructed that executes the proper action at the proper time.

In \cite{pereira2015framework} two types of learning nodes are suggested, the learning action node and the learning composite node. In the learning action node, the user defines a complete RL problem, including states, actions and reward. In the learning composite node, the actions are the children, either leaves or subtrees themselves, but the states as well as the reward are defined by the user.
In a similar spirit,  \cite{fu2016reinforcement} and \cite{zhang2017combining}, replace each fallback node with an RL-node, using its children as actions and user defined reward. The q-values are estimated, and the node executes the child with highest q-value, given the current state.
In \cite{mayr2021learning}, parameters throughout a BT, such as goal points of actions and threshold values of conditions, was learned simultaneously using an evolutionary strategy.

The idea of learning the structure of a BT using genetic algorithms was first proposed in \cite{lim2010evolving}. It was noted how the BT structure, with identical interfaces between subtrees on all levels, provided a policy representation that could easily be subjected to operations such as crossovers and mutations. These ideas were extended in \cite{nicolau2016evolutionary} where grammatical evolution was used to enforce a given structure to the investigated BTs.
Another approach for learning BT structures using genetic algorithms was used in \cite{colledanchise2018learning}. There, the and-or tree analogy of BTs was used to keep the size of the created trees down, while adding new functionality.

The approach proposed in this paper goes beyond the work described above by looking at a particular subset of BTs with a backchained structure \cite{colledanchise2019towards} and using the convergence proof of such BTs in \cite{ogren2020convergence}, to provide insights into how to define individual learning problems that are adapted to the rest of the BT.

\section{Background}
\label{sec:background}
In this section we will give a very brief description of Markov Decision Processes (MDPs), and a more detailed description of BTs, designing backchained BTs and results regarding their convergence.

\subsection{Reinforcement learning}
Reinforcement learning is generally applied to find a control policy $\pi(s) = a$ for a MDP, \cite{howard:dp,sutton2018reinforcement}, defined as follows.

\begin{definition}
\label{def:MDP}
(Markov Decision Process). A MDP is a 4-tuple
\begin{equation}\label{eq:markovDecisionProcess}
(X, A, P_a, R_a ),
\end{equation}
where $X$ is a set of states, $A$ is a set of actions, $P_a(x,x')=P(x_{t+1}=x'|x_t=x,a_t=a)$ is the probability of state transitions,
and
$R_a(x,x')$ is the reward for transitioning to state $x'$ from state $x$. 
\end{definition}

\subsection{Behaviour tree definition}
\begin{definition}[Behavior tree]
\label{def_bt}
 A BT $\bt_i$ is a pair
 
\begin{equation}
 \bt_i = (u_i, r_i)
\end{equation}
where $i$ is an index, $u_i:X \rightarrow U$ is the controller that runs when the BT is executing, and $r_i:X \rightarrow \{\mathcal{R},\mathcal{S},\mathcal{F} \}$ provides metadata regarding the applicability and progress of the execution.
\end{definition}

A BT can either be created through a hierarchical combination of other BTs, using the Sequence and Fallback operators described below, or it can be defined by directly specifying $u_i(x)$ and $r_i(x)$.

The metadata $r_i$ is interpreted as follows: 
\emph{Running} ($\mathcal{R}$),
\emph{Success} ($\mathcal{S}$), and
\emph{Failure} ($\mathcal{F}$).
Let the Running region ($R_i$),
Success region ($S_i$) and
Failure region ($F_i$) correspond to a partitioning of the state space,  defined as follows:
\begin{align}
 R_i&=\{x \in X: r_i(x)=\mathcal{R} \}, \\
 S_i&=\{x\in X: r_i(x)=\mathcal{S} \}, \\
 F_i&=\{x\in X: r_i(x)=\mathcal{F} \}. 
\end{align}

\begin{definition}
\label{def_execution}
Assuming the BT $\bt_i$ is the root, and not a subtree of another BT, and $x\in R_i$,
the system evolves according to $\dot x = f(x,u_i(x))$ or $x_{t+1}=f(x_t,u_i(x))$ depending on if the system is continuous time or discrete time.

\end{definition}

\begin{definition}\emph{(Sequence Compositions of BTs)}
\label{def_sequence}
 Two or more BTs  can be composed into a more complex BT using a Sequence operator,
 $\bt_0=\mbox{Sequence}(\bt_1,\bt_2).$ 
 Then $r_0,u_0$ are defined as follows
\begin{eqnarray}
   \mbox{If }x\in S_1:& %\label{eq:sequence_definition} \\
   &r_0(x) =  r_2(x),~ %\label{bts:eq:seq1a}\\
   u_0(x) =  u_2(x) \label{bts:eq:seq1}\\ 
   \mbox{ else}:& %\nonumber \\
   &r_0(x) =  r_1(x),~% \label{bts:eq:seq2a}\\
   u_0(x) =  u_1(x) \label{bts:eq:seq2}
 \end{eqnarray}
\end{definition}
$\bt_1$ and $\bt_2$ are called children of $\bt_0$. Note that when executing $\bt_0$, the first child $\bt_1$ in (\ref{bts:eq:seq2}) is executed as long as it returns \emph{Running} or \emph{Failure} $(x_k \not \in S_1)$.  
The second child of the Sequence is executed in (\ref{bts:eq:seq1}), only when the first returns \emph{Success} $(x_k \in S_1)$. Finally, the Sequence itself,  $\bt_0$ returns \emph{Success} only when all children have succeeded $(x \in S_1 \cap S_2)$.

For notational convenience, we write
\begin{equation}
\mbox{Sequence}(\bt_1, \mbox{Sequence}(\bt_2,\bt_3))= \mbox{Sequence}(\bt_1,\bt_2, \bt_3),
\end{equation}
and similarly for arbitrarily long compositions. The sequence node is also denoted by ($\rightarrow$), as seen in Figure~\ref{fig:ACC_full_BT}.

\begin{definition}\emph{(Fallback Compositions of BTs)}
\label{def_fallback}
 Two or more BTs  can be composed into a more complex BT using a Fallback operator,
 $\bt_0=\mbox{Fallback}(\bt_1,\bt_2).$ 
 Then $r_0,u_0$ are defined as follows
\begin{align}
   \mbox{If }x\in {F}_1:& %\label{eq:fallback_definition} %\\
   &r_0(x) =  r_2(x),~ %\label{bts:eq:fall1a} %\\
   u_0(x) =  u_2(x)  \label{bts:eq:fall1} \\ 
   \mbox{ else}:& %\nonumber % \\
   &r_0(x) =  r_1(x),~ %\label{bts:eq:fall2a} %\\
   u_0(x) =  u_1(x) \label{bts:eq:fall2}
 \end{align}
\end{definition}

Note that when executing the new BT, $\bt_0$  first keeps executing its first child $\bt_1$, in (\ref{bts:eq:fall2}) as long as it returns  \emph{Running} or \emph{Success} $(x \not \in F_1)$.  
The second child of the Fallback is executed in (\ref{bts:eq:fall1}), only when the first returns \emph{Failure} $(x \in F_1)$. Finally, the Fallback itself $\bt_0$ returns \emph{Failure} only when all children have been tried, but failed $(x \in F_1 \cap F_2)$, hence the name Fallback.

 For notational convenience, we write
\begin{equation}
\mbox{Fallback}(\bt_1, \mbox{Fallback}(\bt_2,\bt_3))= \mbox{Fallback}(\bt_1,\bt_2, \bt_3),
\end{equation}
and similarly for arbitrarily long compositions. The fallback node is also denoted by ($?$), as seen in Figure~\ref{fig:ACC_full_BT}.

\subsection{Backward chained BTs}

In this section we will describe the backward chained approach that was suggested in \cite{colledanchise2019towards}
and parts of the convergence analysis described in \cite{ogren2020convergence}.

\begin{figure}[!h]
    \centering
    \includegraphics[width=\columnwidth]{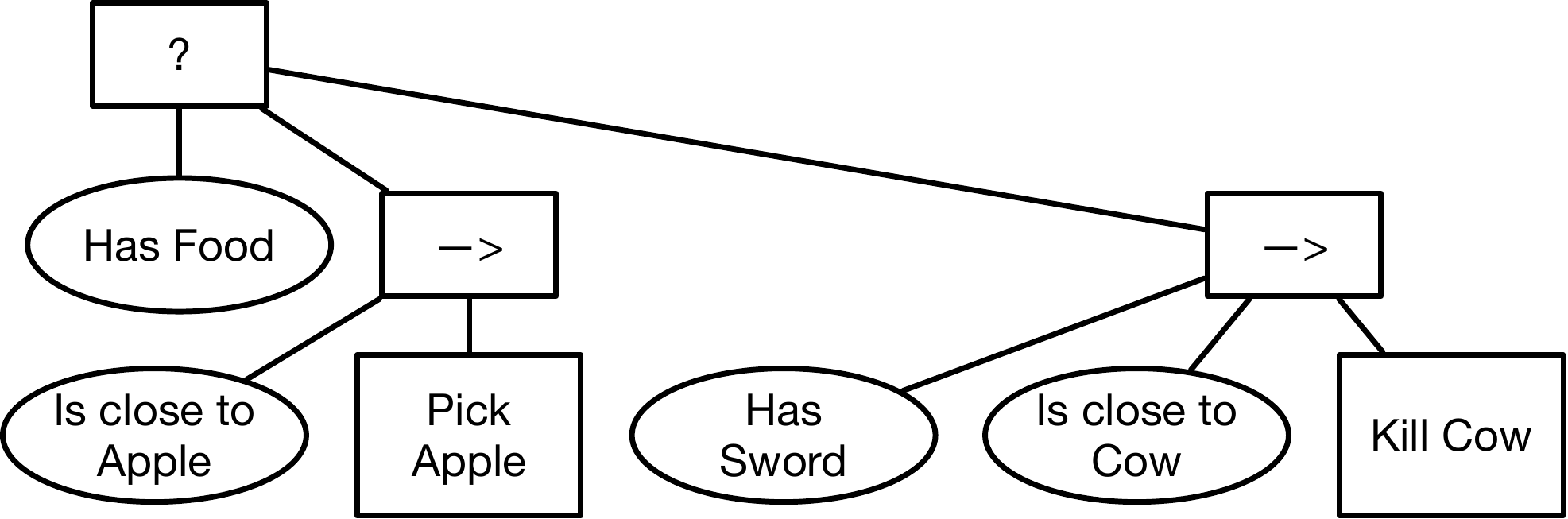}
    \caption{A small BT designed to first check is the agent has food, and if not try to make sure it does.}
    \label{fig:single_PPA}
\end{figure}

Consider the example BT in Figure~\ref{fig:single_PPA}, designed to make sure that the agent has food.
If the agent has food it will immediately return success. If not, it will act to make the condition true.
Sometimes there are multiple ways of making a condition true, in those cases the options are collected under a fallback node, so if one option \emph{(Pick Apple)} fails, or is not applicable, another option \emph{(Kill Cow)} can be invoked. Both actions have their own preconditions, describing when they can be invoked, as illustrated in Figure~\ref{fig:single_PPA}.
Given a list of the available actions, with preconditions and postconditions, such as the one in Table \ref{tab:actions}, one can collect all actions having the same postconditions and create a small BTs of the form shown in Figure~\ref{fig:single_PPA}. 

\begin{table}[!h]
    \centering
    \caption{A list of available actions, with preconditions and postconditions.}
    \label{tab:actions}
    \resizebox{.99\columnwidth}{!}{
    \begin{tabular}{|c|c|c|}
    \hline
        Action & Precondition & Postcondition \\
        \hline
          Escape from fire & - & Safe from fire   \\
        Defeat hostile & - & Safe from hostiles  \\
         Eat      & Has food& Not hungry  \\
        Pick Apple & Is close to apple & Has food  \\
        Kill Cow & Is close to cow, has sword & Has food   \\
        Craft sword & Has materials, Has crafting table & Has sword \\
         Chase cow & Can see cow & Close to cow  \\
         Search for cow & -  &  Can see cow\\
        \vdots & \vdots & \vdots  \\
        \hline
    \end{tabular}
    }
\end{table}

\begin{figure}[!t]
    \centering
    \includegraphics[width=\columnwidth]{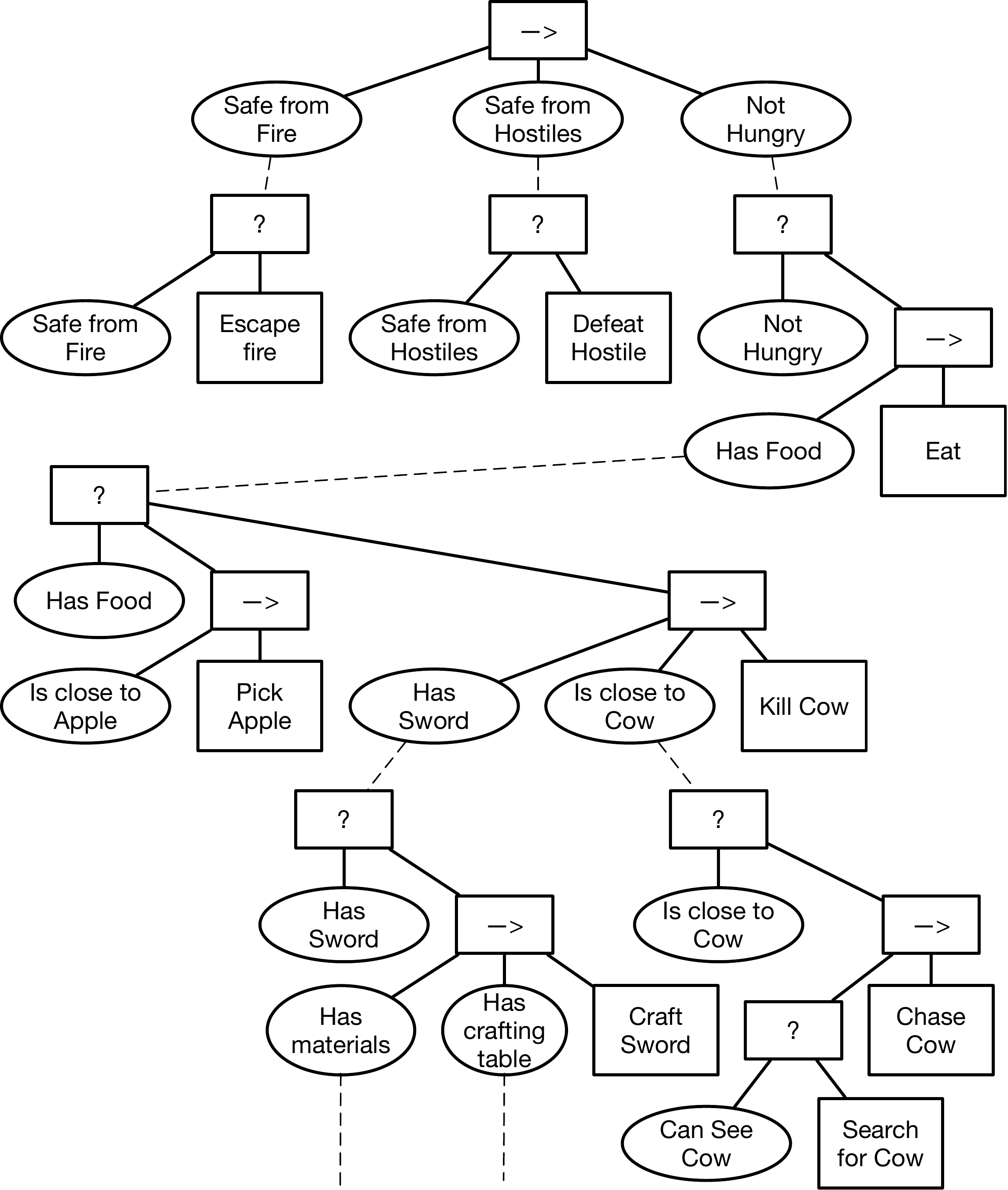}
    \caption{A backward chained BT can be created by recursively replacing (along dashed lines) condition checks with BTs trying to achieve those conditions. The process is started with the three top level goals listed at the top of the figure. The result can be seen in Figure~\ref{fig:ACC_full_BT}.}
    \label{fig:PPA_connect}
\end{figure}

The key idea is now to recursively apply the design in Figure~\ref{fig:single_PPA}, as illustrated in Figure~\ref{fig:PPA_connect}.
First we list three top priority goals in a sequence node, safe from fire, safe from hostiles and not hungry. Then, 
instead of just checking the conditions we can replace them (illustrated by dashed lines)  with a small BT, of the form shown in Figure~\ref{fig:single_PPA}, to make them true.
The resulting BT will have some new conditions, which in turn can be replaced by small BTs achieving them, and so on. This process is formally described in Algorithm \ref{alg:backward_chained}.

\begin{algorithm2e}[h]
\KwIn{ Goals $C_1,\ldots, C_M$}

$\bt_0 \gets$ \FuncSty{Sequence $(C_1,\ldots, C_M)$}\;

  \While{Exists $C' \in \bt_0$ such that $\FuncSty{Parent}(C') = \FuncSty{Sequence}$ 
  AND
  $C'$ is postcondition of some action(s) $A_j$}{
  $\bt'\gets$  \FuncSty{Fallb($C_i$, Seq(Pre($A_j$),$A_j$), $\ldots$)}\;
 \FuncSty{Replace $C_i$ in $\bt_0$ with $\bt'$ }
    }
\caption{Creating a Backward Chained BT}
\label{alg:backward_chained}
\end{algorithm2e}

The class of BTs we will analyse is defined as follows.

\begin{definition}\emph{(Backward chained BT)}
A  Backward chained BT is a BT that is constructed from a set of desired top level goal conditions in a Sequence, that are then recursively replaced by BTs achieving them, as described in Algorithm~\ref{alg:backward_chained}.
\end{definition}

In \cite{ogren2020convergence}, sufficient conditions for convergence of backchained BTs were presented. A key part of the requirements were that each action does not violate its own ACCs, defined as follows.

\begin{definition}[Active Constraint Conditions (ACC)]
\label{def_ACC}
Given a BT $\bt_0$, and an action $A_i$ in that BT, the Active Constraint Conditions $ACC(i)$ of $A_i$ is the sets of conditions,  apart from the preconditions of $A$,  that needs to return \emph{Success} (be true) for $A$ to execute.
\end{definition}

Looking at Figure \ref{fig:ACC_full_BT}, we see that the ACCs of \emph{Chase Cow} has three ACCs: \emph{Safe from Fire}, \emph{Safe from Hostiles} and \emph{Has Sword}. Note that if one of these is violated, the agent has to switch to another action, to satisfy the condition again. If the agent comes too close to fire, it will invoke \emph{Escape fire}. If the agent is threatened by a hostile entity, it will invoke \emph{Defeat Hostile}, and if it brakes the sword it will have to make a new one. Thus, violating an ACC once might delay progress to achieving all goals, while repeated violation of ACCs might result in infinite loops where the agent is stuck undoing its own actions. To illustrate the latter case we note that if \emph{Chase Cow} is completely unaware of fires, and there happens to be a fire between the agent and the cow, then the agent will move towards the cow, thereby getting too close to the fire, switch to \emph{Escape fire}, thereby moving away from the cow, switch back to \emph{Chase Cow} and again move towards the fire, and so on indefinitely.

Note that ACCs is not simply a list of all achieved subgoals. In order to craft the sword, the agent had to have materials and a crafting table. But once it has the sword, those subgoals are no longer important. Thus ACCs are exactly the important subgoals that should not be violated.

The key idea of this paper is to use the information embedded in the ACC concept, taken from the proof in \cite{ogren2020convergence}, in the definition of RL problems for individual actions.

\section{Proposed approach}
\label{sec:proposed_approach}
In this section we will first state the problem we are trying to solve, and then provide the details of the solution we are proposing.

\subsection{Problem}
The problem we are addressing is what to do when a backchained BT created using Algorithm \ref{alg:backward_chained}, that includes RL policies for some or all of the actions does not succeed, i.e., it does not end up in a state where the overall BT returns success ($x \not \in S_0$). Instead, it might end up in a state where the BT returns failure ($x  \in F_0$), or it keeps running indefinitely ($x \in R_0$), either executing the same action or switching between a set of actions.

Looking at the example in Figure~\ref{fig:ACC_full_BT} we could either have that \emph{Chase Cow} is never able to catch up with the cow, or that there are some hostile agents around and when chasing the cow the agent comes too close to the hostiles, resulting in a switch to \emph{Defeat hostile}, followed by a switch back to \emph{Chase Cow} and so on. As we will see below, this problem can be addressed in a structured way of making the agent chase the cow but at the same time trying to stay away from hostiles.

We assume that the world the agent inhabits can be modelled by a (MDP), see Definition~\ref{def:MDP}. 

\subsection{Creating an action using RL}\label{sec:RLaction}
We propose to address the problem above by improving the performance of the BT using a combination of RL and ideas from the convergence proof described in \cite{ogren2020convergence}.
Formulating the RL problem, we will not only use the desired postcondition (see Table \ref{tab:actions}) as a positive reward, but also take ACCs (Definition \ref{def_ACC}) into account by applying a negative reward and/or ending the episode when an ACC is violated.

\begin{definition}[RL problem for creating an Action $\bt_i$]
\label{def:RL_problem}
Let the reward be given by three constants $M_p >> 0 > M_t > M_{ACC}$, with p for postcondition and t for time, as
\begin{align}
    R_{a}(x',x) 
    &= M_p >> 0 &\mbox{ if } x' \in  S_i\\
    &= M_{ACC} < 0 &\mbox{ if } x' \not \in \bigcap_{j \in ACC(i)} S_j \\
    &= M_t < 0 &\mbox{ else}. 
\end{align}
The episode is ended when the postcondition is achieved, the agent dies, or the maximum number of time steps is reached. We will additionally evaluate the effect of also ending the episode when an ACC is violated. 
After training we let $\bt_i$ be given in terms of the learned state-value function $q_i(x,a)$ as described below. 

\begin{align}
    u_{i}(x) &= \mbox{argmax}_a q_i(x,a) 
\end{align}
\end{definition}

The reason for ending the episode upon ACC violation is as follows:
During the normal execution of the BT, once the policy is trained, a violation of an ACC would by definition cause the BT to switch to a different action. Thereby it has the same effect as ending the episode - the action is forced to stop executing in favour of another action. Similarly, in this version of training, the episode is ended, but the environment is not reset. The training resumes with a new episode after the ACC violation has been resolved by the rest of the tree. This results in episodes that better represent the situations that the policy will encounter during real execution. Additionally, it stops the agent from accumulating further reward during the episode, making the impact of a single ACC violation larger. 

\subsection{When to apply RL to actions}
As stated above, the main problem we are addressing in this letter is to improve the performance of an existing backward chained BT. Either because it reaches a state where it returns failure, or because it executes forever, possibly looping through the same set of actions, or just keeps executing a single one.

In light of the convergence proof of \cite{ogren2020convergence}, one would then try to identify actions that violate their ACCs or fail to reach their their postconditions, and then apply RL to them using Definition \ref{def:RL_problem}.

Looking at Figure \ref{fig:ACC_full_BT}, we see that an example of failure to reach the postcondition might be
that \emph{Escape fire} does not get the agent out of the fire.
Furthermore, examples of ACC violations might be that \emph{Defeat hostile} keeps running into the fire and thereby initiates the execution of \emph{Escape fire}, or that \emph{Chase cow} moves too close to a hostile agent and thereby initiates the execution of \emph{Defeat hostile}.

Beyond improving the performance of a BT with existing controllers, one could also use Definition \ref{def:RL_problem} to learn all controllers of a BT.
Imagine a BT that was created using Algorithm~\ref{alg:backward_chained} from a list of action names with  preconditions and postconditions, such as Table~\ref{tab:actions}.
Then we have no implementations beyond the conditions, but a clear description of when to run each action, and what postconditions it should achieve and what ACCs it should not violate.
Given these, we can apply the RL problem in Definition~\ref{def:RL_problem} to find implementations, $u_i$, for all actions.

Note that Definition~\ref{def:RL_problem} is different from the baseline approach of just running the RL based on the information in Table~\ref{tab:actions}, i.e., only trying to achieve the postconditions, without taking the ACCs into account.
In the next section we will compare the proposed approach with such a baseline method.

\section{Experimental results}
\label{sec:experimental_results}
In this section we will provide an empirical analysis of the method proposed above, identifying scenarios when it has significant impact on performance and scenarios when it has not. We first give an overview of the technical setup of the environment and experimental configurations and then present the results of our experiments.
As will be seen, the benefits of taking ACCs into account varies with the context. If violating the earlier achieved objective can be remedied quickly (such as stepping away from the fire), and there is no risk of getting stuck in a pattern of repeatedly doing and undoing the same action, then the performance improvements will be very small. If however, it takes considerable  time to achieve the objective (such as defeating a hostile agent), or there is risk of infinite loops, then the performance improvement can be significant. The experiments below will show examples of both these cases.

\subsection{Technical setup}
The technical setup for our experiments is based upon the simulation environment Project Malmö \cite{malmoplatform} and the RL algorithms in Stable-Baselines 3, \cite{stable-baselines3}. These modules were chosen because they provide an extremely dynamic environment with countless objects and other agents that can be interacted with in numerous ways, while at the same time being freely accessible enabling reproducible results. 
The source code of the resulting technical solution is available online at \cite{ourrepository} alongside the results of the training.

Project Malmö \cite{malmoplatform} is an open-source platform developed by Microsoft Research, for the purpose of supporting AI experiments by providing a simulation environment based on the Minecraft engine. Malmö is a customisable virtual environment which allows for the testing of algorithms on tasks freely determined by the users. The Malmö interface allows for custom design of so called missions, which can be used to recreate common AI and Robotics related problem scenarios ranging from path-planning to multi-agent collaboration. One of the technical aspects of Malmö is that the control of the agent happens asynchronously from the stepping of the environment. This sets time constraints on the decision making process for an agent acting in the environment, in a way that is similar to a non-simulated system.

For RL we decided to use the open-source implementation Stable-Baselines 3 \cite{stable-baselines3}.
It is based on Open-AI Gym \cite{brockman2016openai} and Baselines \cite{baselines}, and provides a customisable Gym interface which can be used to integrate it with many different environments, including  Project Malmö. The central Gym interface allows for integration to different environments as well as switching between different RL algorithms, as long as they implement the Gym interface. At the time of writing, Stable-Baselines supports many widely known RL algorithms, such as PPO \cite{schulman2017proximal}, DQN \cite{DQN} and A2C \cite{A2C}. We used PPO for the experiments of this letter.

\subsection{Experimental setup}\label{sec:expsetup}
The primary aim of the experiments is to investigate to what extent the setup in Definition \ref{def:RL_problem} does improve the efficiency of the BT execution, by not violating the previously achieved subgoals, as encoded by the ACCs.

We define our experiments based on the agent in Figure~\ref{fig:ACC_full_BT}.
Two different actions, \emph{Defeat hostile} and \emph{Chase cow}, will be trained using the RL problem in Definition \ref{def:RL_problem}, and the complete BT will be evaluated in two different scenarios, see Table \ref{tab:scenarios}, one starting near a hostile, away from fire and not hungry,  and the other starting hungry, but currently away from fire and hostiles.

\begin{table}[!h]
    \centering
    \caption{The two scenarios to be evaluated, with remaining top level goal emphasized.} 
    \label{tab:scenarios}
        \begin{tabular}{|c||c|}
        \hline
             &  
            \textbf{Starting state}  \\
            \hline \hline
            Scenario 1           & Safe from fire,  \textbf{Not Safe from Hostiles}, Not Hungry       \\
            Scenario 2           & Safe from fire, Safe from Hostiles,   \textbf{Hungry}       \\
            \hline
        \end{tabular}
\end{table}

We will evaluate four different designs, see Table~\ref{tab:configurations}. First one where the two actions are trained using a standard RL reward, taking only the desired postcondition into account.
Then three different variations of ACC-aware RL, following Definition \ref{def:RL_problem}.
There are two ways to take the ACC into account. First using a negative reward $M_{ACC}$, and second by ending the episode when the ACC is violated (this is what happens when the BT starts executing another action). 
We will evaluate all tree combinations of these two options, see Table~\ref{tab:configurations}.

\begin{table}[!h]
    \centering
    \caption{Parameters of RL problems} 
    \label{tab:configurations}
        \begin{tabular}{|p{31mm}||p{18mm}|p{22mm}|}
        \hline
            \textbf{Configuration} & %\textbf{Reward for achieving post\-condition} & 
            \textbf{Reward for violating ACC $M_{ACC}$} & \textbf{End episode when violating ACC} \\
            \hline \hline
            Standard RL             & 0         & False   \\
            ACC Aware (Neg. Reward) & -10       & False   \\
            ACC Aware (End Episode) & 0         & True    \\
            ACC Aware (NR and EE)   & -1000     & True    \\
            \hline
            \multicolumn{3}{l}{\emph{All configurations use $M_p=1000$ and $M_t=-0.1$, see Definition \ref{def:RL_problem}. }}
        \end{tabular}
\end{table}

As mentioned above, we train policies for two different BT actions,
\emph{Defeat hostile} and \emph{Chase cow}.
In Scenario 1, see Table~\ref{tab:scenarios}, the agent is not hungry, so \emph{Chase cow} will not be executed, but in Scenario 2 both the trained actions might be used.

 \begin{table}[!h]
    \centering
    \caption{ACCs of the two to be trained.} 
    \label{tab:acc}
        \begin{tabular}{|c||c|}
        \hline
             &  
            \textbf{ACCs}  \\
            \hline \hline
            Defeat hostiles           & Safe from fire \\
            Chase cow          & Safe from fire, Safe from Hostiles, Has sword       \\
            \hline
        \end{tabular}
\end{table}

The ACCs of the two actions to be trained can be seen in Table~\ref{tab:acc}. Thus, when defeating hostiles the agent should try to keep out of the fire, while when chasing the cow it should try to keep out of the fire, keep out of the way of hostiles, and avoid losing the sword.

\subsection{Training Results}
\begin{figure}[!h]
    \centering
    \includegraphics[width=\columnwidth]{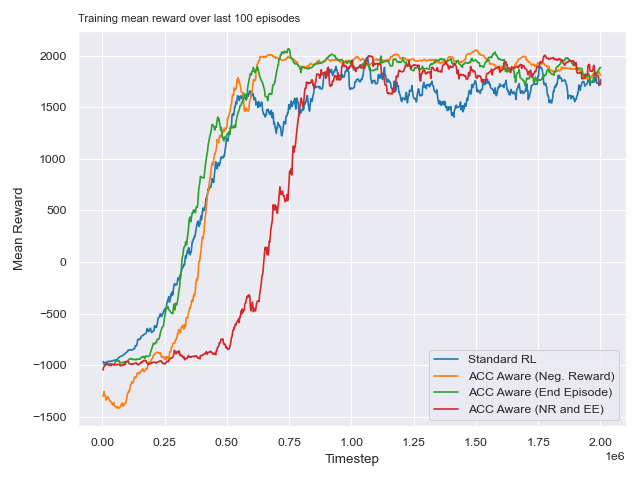}
    \caption{Training results for Defeat hostile.}
    \label{fig:rewards_s}
\end{figure}

\begin{figure}[!h]
% Needs an update!!!
    \centering
    \includegraphics[width=\columnwidth]{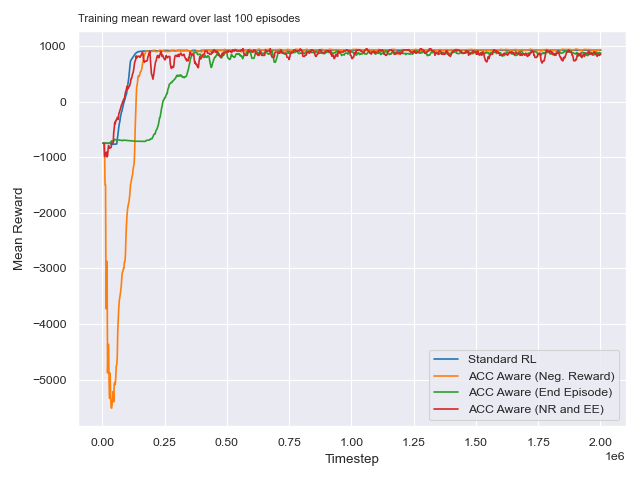} 
    \caption{Training results for Chase cow.}
    \label{fig:rewards_c}
\end{figure}

The PPO algorithm \cite{schulman2017proximal} was used for training in all cases. Both the policy and the value function were modelled as a Multi-Layer Perceptron, with default parameters from Stable-Baselines used during the training process. Each RL policy was trained for 2 million timesteps, using the configurations described in Table \ref{tab:configurations}. The training results are shown in Figures \ref{fig:rewards_s} and \ref{fig:rewards_c}. 

It can be seen that the training has converged to policies that 
consistently produce good rewards in the all of the respective configurations.
There is a larger variance of rewards in the later stages of training for the \emph{Defeat hostile} policies in Figure \ref{fig:rewards_s}. This is caused by the fact that defeating hostiles is a very dynamic activity that can play out in many different ways.
 Even a small mistake can lead the agent to be pushed into the fire by the opposing agent. As the training will still keep experimenting with small random deviations from the currently known best policy in the later training stages, it can lead to relatively large deviations in accumulated reward during training, as small changes in behavior can quickly lead to large negative rewards. 
There is a smaller variance in \emph{Chase cow}, as the hostile is mostly static while not having seen the agent.

\subsection{Evaluation Results}
To see the effect on the performance of the overall BTs,
 including trained actions, were ran evaluations for 1000 episodes in the two scenarios described in Table~\ref{tab:scenarios}.

The numerical evaluation metrics, as seen in Tables~\ref{tab:evalfighter} and \ref{tab:evalchaser}, were the time taken for mission completion, and the number of ACC violations both as a percentage of episodes with at least one violation and as the average number of ACC violations per episode.

Looking at the data from Scenario 1 in Table~\ref{tab:evalfighter}, we see that the differences in completion time are fairly small (less than 10\%), with the \emph{ACC-aware (Neg. Reward)} configuration performing best, but only slightly better than the \emph{ACC-aware (End Episode)} followed by \emph{standard RL}. The worst performing is the \emph{ACC-aware (NR and EE)}.
We can also see that for the \emph{standard RL} configuration the percentage of episodes with at least one ACC violation are fairly high (21\%), but in each episode it only spends less than 0.5\% of the total number of time steps (2.1 time steps out of 446)  violating those ACCs.
Thus, in Scenario 1 we see no clear performance improvement from using the ACCs. 

The reason for these outcomes is that the ACC violation of \emph{Safe from fire} can be fixed in a very short amount of time by moving out of the fire. Therefore, even though \emph{standard RL} does violate an ACC in many episodes, the time spent violating the ACC is fairly small, and the resulting impact on average completion time is also fairly small.
If, however, there would have been cases of infinite loops of violating and achieving an ACC, an easily fixed violation would have had a larger impact on performance.

\begin{table}[]
    \centering
    \caption{Evaluation of Scenario 1, starting not hungry but close to hostile.}
    \label{tab:evalfighter}
    \begin{tabular}{|p{18mm}||p{17mm}|p{8mm}|p{5mm}|p{7mm}|p{5mm}|}
        \hline
        \textbf{Configuration} & ACC* violations (\% episodes) &
       %  ACC* violations (avg. \# steps) & 
                  \multicolumn{2}{p{17mm}|}{ ACC* violations (\# steps)}   &
         \multicolumn{2}{p{18mm}|}{ Time (steps) to completion}   \\ 
         \hline
           & \% & Mean & SD & Mean & SD \\
        \hline
         \hline
        Standard RL                %   & 17\% & 1.9 &5.83 & 444.5 & 5.83 \\ 
         & 20.8\% & 2.1 & 5.24  & 446 & 78 \\ 
        \hline
        ACC aware RL (Neg. Reward)    %& 6\%  & 0.36 && \textbf{431.9} & 2.00  \\
      %  Tex string: & 5.6000000000000005 & 0.407 & 1.9640140019867474 & 0.9584083292749999 & 0.6 & 422.342 & 76.60770872438361 
 & 5.60\% & 0.4 & 1.96  & \textbf{422} & 77 \\ 
        \hline
        ACC aware RL (End Episode)  %  & 3\%  & \textbf{0.12} && 437.7 &1.77 \\
       % Tex string: & 3.4000000000000004 & 0.235 & 1.386280996046617 & 0.9569083269 & 0.6 & 437.289 & 85.34653759233588 
 & 3.40\% & 0.24 & 1.39  &  437 & 85 \\ 

        \hline
        ACC aware RL (NR and EE)    %  & \textbf{2\%}  & 0.18 && 474.7 &1.10 \\
     %   Tex string: & 2.9000000000000004 & 0.154 & 0.9382345122622595 & 0.9567083268500001 & 0.2 & 460.175 & 76.35293298230265 
 & \textbf{2.90\%} & \textbf{ 0.15} & 0.94  & 460 & 76 \\
        \hline
        \multicolumn{4}{l}{\emph{*ACC = "Safe from Fire"}}
    \end{tabular}
\end{table}

\begin{table}[]
    \centering
    \caption{Evaluation of Scenario 2, starting hungry.}
    \label{tab:evalchaser}
     \begin{tabular}{|p{18mm}||p{17mm}|p{8mm}|p{5mm}|p{7mm}|p{5mm}|}
        \hline
        \textbf{Configuration} & ACC* violations (\% episodes) &
       %  ACC* violations (avg. \# steps) & 
                  \multicolumn{2}{p{17mm}|}{ ACC* violations (\# steps)}   &
         \multicolumn{2}{p{18mm}|}{ Time (steps) to completion}   \\ 
         \hline
          & \% & Mean & SD & Mean & SD \\
        \hline
         \hline
        Standard RL                %   & 17\% & 1.9 &5.83 & 444.5 & 5.83 \\ 
       % Tex string: & 100.0 & 790.163 & 89.57436257657656 & 1.0 & 0.0 & 952.308 & 89.3640371514179 
& 100.0\% & 790 & 90  & 952 & 89  \\
        \hline
        ACC aware RL (Neg. Reward)    %& 6\%  & 0.36 && \textbf{431.9} & 2.00  \\
%Tex string: & 18.6 & 133.661 & 300.1734233389091 & 1.0 & 0.0 & 303.087 & 301.3913592507257 
& 18.6\% & 134 & 300  & 303& 301 \\
        \hline
        ACC aware RL (End Episode)  %  & 3\%  & \textbf{0.12} && 437.7 &1.77 \\
%Tex string: & 1.0999999999999999 & 6.7 & 70.86694575046958 & 1.0 & 0.0 & 183.833 & 71.27642745676862 
 & 1.1\% & 6.7 & 71 & 184 & 71 \\ 

        \hline
        ACC aware RL (NR and EE)    %  & \textbf{2\%}  & 0.18 && 474.7 &1.10 \\
%Tex string: & 0.0 & 0.0 & 0.0 & 1.0 & 0.0 & 176.286 & 7.238660373301126 
& \textbf{0.0\%}  &\textbf{0.0}   & 0.0  & \textbf{176}  & 7.2 \\
        \hline
        \multicolumn{4}{l}{\emph{*ACC = "Safe from Fire"}}
    \end{tabular}

\end{table}

Looking at the data from Scenario 2 in Table~\ref{tab:evalchaser}, we see that the differences in completion time are fairly large (more than 500\%), with the \emph{ACC-aware (NR and EE)} configuration performing best, followed by \emph{ACC-aware (End Episode)} and \emph{ACC-aware (Neg. Reward)}.
We can also see that for the \emph{standard RL} configuration the percentage of episodes with at least one ACC violation is very high (100\%), and in each episode over 50\% of the total number of time steps (790 time steps out of 952)  violate ACCs.
Thus, in Scenario 2 we see a very clear performance improvement from using the ACCs. 

The reason for these outcomes is that the ACC violation of \emph{Safe from hostiles} leads to the hostile agent approaching the agent and a subsequent fight. So even though all agents are able to eventually defeat the hostile, the fight itself creates a significant delay in reaching the cow. 

Looking at the difference between the three ACC-aware configurations in Tables~\ref{tab:evalfighter} and \ref{tab:evalchaser} we note the following.
The negative reward alone has an effect on the ACC violation rate, but on average has a smaller impact than  ending the episode. It could be argued that larger negative rewards might have a larger impact, but it could theoretically lead to instability if it is too large, as this reward would be given for every timestep in which the ACC is violated. 

The "End Episode" configuration has a strong effect on ACC violations in both scenarios. This sort of ending to the episode is always a negative occurrence, because the potential positive reward can no-longer be reached, leaving the agent with a lower cumulative reward. 
Using the configuration of "End Episode" and "Negative Reward" together appears to have the largest overall effect. This option not only prevents future progress towards a positive reward, but also lowers the reward accumulated during the episode, without the risk of instability as the large reward is only given once.

To summarise, we note that avoiding ACC violations can have either a very small or a very large impact on the overall performance of the BT, in a way that depends on how much extra effort is needed to re-achieve the ACC. Furthermore, the strongest effect in terms of reducing the ACC violations is given by the combination of a negative reward and ending the episode.

\section{Conclusions}
\label{sec:conclusions}
This letter investigated the use of RL policies for improving the performance of backward chained BTs.
The key idea was to use the ACCs, that were defined in the literature as an important part of theoretical convergence proofs,
in the setup of the RL problems for individual actions.
Our experiments show that the proposed approach can produce significant performance improvements in cases where achieving the ACC takes significant effort (such as defeating a hostile). The improvements can be smaller in cases where achieving the ACC requires less effort (stepping away from fire) or there is no risk of getting caught in a loop of repeatedly doing and undoing the same action. 

Finally, we conclude that ACCs can be used to both analyse the BT, when considering the implementation of an RL based action, and to provide inputs to the reward function and the training process itself.

\vspace{12pt}
\bibliographystyle{IEEEtran}
\bibliography{references}
\end{document}